\documentclass[prl,aps,showpacs,twocolumn]{revtex4-1}
\usepackage{times}
\usepackage{amssymb}
\usepackage{amsmath}
\usepackage{mathrsfs}
\usepackage{graphicx}
\usepackage{epsfig}
\usepackage{dcolumn}
\usepackage{color}
\usepackage{bm}
\usepackage[table,xcdraw]{xcolor}
\usepackage{cleveref}
\usepackage{xr}

\begin{document}

\title{Multi-partite entanglement detection with non symmetric probing}
\author{Luca Dellantonio, Sumanta Das, J\"{u}rgen Appel, and Anders S. S\o rensen}
\affiliation{The Niels Bohr Institute, University of Copenhagen, Blegdamsvej 17, DK-2100 Copenhagen \O, Denmark\\}
\date{\today}

\begin{abstract}
%We propose a novel technique that evaluates the squeezing of quantum fluctuations in an atomic ensemble for an inhomogeneous laser-atom interaction. 
We show that spin squeezing criteria commonly used for entanglement detection can be erroneous, if the probe is not symmetric. We then derive a lower bound on squeezing for separable states in spin systems probed asymmetrically. Using this we further develop a procedure that allows us to verify the degree of entanglement of a quantum state in the spin system. Finally, we apply our method for entanglement verification to existing experimental data, and use it to prove the existence of tri-partite entanglement in a spin squeezed atomic ensemble.
\end{abstract}

\maketitle

Entanglement is a fundamental resource for proving non-classicality of nature, and is a key feature for developing quantum technologies. It was first used experimentally for proving the breakdown of local realism \cite{Bell,Aspect}, and has now become a practical ingredient in quantum information science and metrology \cite{NielsenChuang}. As such, it is crucial to discern between separable and entangled systems. Unfortunately, this is not always straightforward; even though several criteria exist \cite{EntMeasure1,EntMeasure2,EntMeasureUGO}, it may be hard to experimentally verify that a state is entangled. Hence, there is a need for simple, practical procedures for proving that a system is entangled. An example of such a criterion is spin squeezing \cite{Weineland,Weineland2,SqueeMain}, which has often been used to probe entanglement in multi-particle systems \cite{Smerzone}. One of its major advantages is that it relies on measuring only two observables: the mean spin and the fluctuations perpendicular to it. This makes it ideally suited for systems where complete control over all degrees of freedom is hard to achieve. Measuring that the noise is squeezed below a certain bound is then a sufficient criterion for proving entanglement \cite{AndersNat,Anders,Vitagliano,Squuez6,Squuez7,EntSque1,EntSque2,EntSque3,EntSque4}. Due to the simplicity of this approach, it has been employed to show even multi-partite entanglement in a wide range of experiments \cite{Squuez1,Squuez2,Squuez5,Squuez4,Thompson}. However, an inherent assumption in the entanglement criteria based on spin squeezing is that all particles are probed with equal strength. In many practical situations this is a very good approximation \cite{Squuez2,Squuez4}, but in others this is far from reality. As an example, consider a Gaussian beam probing a collection of trapped atoms. If the waist of the laser beam is much smaller than the size of the cloud, the atoms will have an asymmetric interaction with the light. This asymmetry leads to minor modifications of the interaction dynamics if suitable weighted operators are introduced \cite{Vuletic,QND6}. For entanglement detection in the ensemble, however, the effect of such asymmetry may be much more severe and has so far not been investigated.

In this article, we consider the effect of asymmetric probing on the entanglement criteria. We first consider a simple generalization of the standard squeezing criterion and show that it is no longer a suitable method for verifying entanglement. To overcome this problem we develop new entanglement criteria, that can accommodate the asymmetric probing of the particles. We show that our criteria are sufficient for detection of bi-partite entanglement as well as higher order multi-partite entanglement. Finally, we apply these criteria to a recent experiment \cite{Appel} to show the existence of tri-partite entanglement in an atomic ensemble of cold, asymmetrically probed Cs atoms.

In spin systems, the collective spin operator $\vec{J}=\sum_{i} \vec{j_{i}}$ is typically used to represent the observables of the system. Here, operators $\vec{j}_{i}$ are the standard (pseudo-)spin operators used to describe the individual two level systems with states $\lvert \uparrow\rangle_{i}$ and $\lvert \downarrow\rangle_{i}$, (eigenstates of $j_{z,i}$). In reality, however, the probing may differ significantly from particle to particle, so that the experimental setup rather measures the \textit{weighted spin operator}
\begin{equation}
\vec{S} = \sum\limits_{i=1}^{N}\eta_{i}\vec{j_{i}},\label{WSO}
\end{equation}
which accounts for the different probing strengths $\eta_{i}$ of the $N$ particles.

To characterize spin squeezing, Wineland \textit{et al.} \cite{Weineland} introduced the squeezing parameter $\xi^2$, which describes the improvement in spectroscopic resolution compared to using coherent spin states $\lvert\text{CSS}\rangle = \bigotimes_{i=1}^{N}\lvert \uparrow\rangle_{i}$. Generalizing the derivation of Ref. \cite{Weineland}, we find that in the case of asymmetric coupling, the parameter $\xi^2$ takes the form
\begin{equation}
\xi^{2}_\text{A} = \frac{(\Delta S_{x})^{2}}{\langle S_{z}\rangle^{2}}\left[ \frac{(\Delta S_{x})^{2}_\text{CSS}}{\langle S_{z}\rangle^{2}_\text{CSS}}\right]^{-1},\label{SP}
\end{equation}
where averages marked with ``CSS'' refer to the state $\lvert \text{CSS}\rangle$, and the subscript ``A'' stands for asymmetric. The standard form $\xi^{2}=\left[\langle J_{z}\rangle^{2}_\text{CSS}/(\Delta J_{x})^{2}_\text{CSS}\right]/\left[\langle J_{z}\rangle^{2}/(\Delta J_{x})^{2}\right]=N\left[(\Delta J_{x})^{2}/\langle J_{z}\rangle^{2}\right]$ of the squeezing parameter is found by setting all coefficients $\eta_{i}=1$ in Eq. \eqref{WSO}, so that $\vec{S}\rightarrow\vec{J}$. Under this constraint, $\xi_{A}^{2}<1$ is a sufficient condition for proving entanglement in the ensemble \cite{AndersNat}. However, if the $\eta_{i}$ are not identical, the situation is different; to show this, we consider the separable state 
\begin{equation}
\lvert \psi \rangle = \prod_{i=1}^{N} \lvert \theta_{i}\rangle,\label{ExSepSt}
\end{equation}
where $\lvert \theta_{i}\rangle$ are single particle states defined by \mbox{$\lvert \theta_{i}\rangle\equiv e^{i\theta_{i}J_{y}} \lvert \uparrow\rangle$}. The average value of the spin along the $z$ direction and the variance along $x$ are then given by $\langle S_{z}\rangle = \sum_{i}\eta_{i}\langle j_{i,z} \rangle$ and $\left(\Delta S_{x}\right)^{2}= \sum_{i}\eta^{2}_{i}\langle j_{i,z} \rangle^{2}$ respectively. Employing the definition of the CSS, we can derive the associated values to be $\langle S_{z}\rangle_{\text{CSS}}=\sum_{i}\eta_{i}/2$ and $ \left(\Delta S_{x}\right)^{2}_{\text{CSS}}=\sum_{i}\eta_{i}^{2}/4 $, so that we can express the asymmetric squeezing parameter in terms of the mean spins and the $\eta_{i}$'s:
\begin{equation}
	\xi_{A}^{2}\left( \lbrace\eta_{i}\rbrace_{i}; \lbrace\langle j_{i,z} \rangle\rbrace_{i} \right)
	=\frac{\left( \sum_{i}\eta_{i}\right) ^{2}}{\sum_{i}\eta_{i}^{2}}\frac{\sum_{i}\eta_{i}^{2}\langle j_{i,z} \rangle^{2}}{\left( \sum_{i}\eta_{i}\langle j_{i,z} \rangle\right)^{2}}.\label{GenSqP}
\end{equation}
Now consider the distribution $\eta_{1}=1$, $\eta_{i}=\epsilon \ll 1$ for all $i= 2,...,N$ and take a quantum state $\lvert\psi\rangle$ such that $\eta_{i}\langle j_{i,z}\rangle=\epsilon$ for all $i$. Then, using Eq. \eqref{GenSqP} we get that to lowest order in $\epsilon$, $\xi^{2}_{A}=\frac{1}{N}+o(\epsilon)$, meaning that there exists a separable state for which $\xi^{2}_{A}$ equals $1/N$ up to an infinitesimal quantity. Moreover, this value can be shown to be the lowest possible for a finite number of particles $N$. As a consequence of the fact that $\xi^{2}_{A} \leq 1$ for a separable state, we conclude that the simple squeezing parameter in Eq. \eqref{SP} cannot be used for verifying entanglement with asymmetric probing. In order to verify entanglement in experiments, it is therefore essential to derive new entanglement measures which are applicable to the actual experimental setups.

In the following we develop new entanglement criteria for spin systems probed asymmetrically. For this purpose we use the Lagrange multiplier method of Ref. \cite{Anders}. We are interested in minimizing the variance of the spin along $x$ for a given value of the mean spin along $z$. We therefore minimize the Lagrange function $\Gamma(\mu)\equiv (\Delta S_{x})^{2} -\mu \langle S_{z}\rangle$, where $\mu$ is a Lagrange multiplier. In this way a condition on entanglement can be established by finding a lower bound on $(\Delta S_{x})^{2}$ for a given value of $\langle S_{z}\rangle$ for a separable state. To do this, the expectation values of the operators $S_{x}, S^{2}_{x}$, and $S_{z}$ are evaluated for a separable state density matrix defined by $\rho = \sum_{k}p_{k}\rho^{(k)}_{1}\otimes...\otimes\rho^{(k)}_{N}$,
where the $p_{k}$'s are positive numbers such that $\sum_{k}p_{k}=1$. In addition, for any $k$ and $i=1,...,N$, $\rho^{(k)}_{i}$ is the density matrix of the $i$-th particle. Using this $\rho$, it is possible to derive the average value for the spin $\langle S_{z}\rangle = \sum_{k}p_{k}\sum_{i=1}^{N}\eta_{i}\langle j_{i,z}\rangle_{k}$ and a lower bound for the variance $\left(\Delta S_{x}\right)^{2}\geq \sum_{k}p_{k}\sum_{i=1}^{N}\eta_{i}^{2}\langle j_{i,z}\rangle_{k}^{2}$. For the latter, we have used $\langle j_{i,{x}}^{2}\rangle - \langle j_{i,{x}}\rangle^{2} \geq \langle j_{i,z}\rangle^{2}$ for spin $\frac{1}{2}$ particles and Jensen's lemma \cite{Jensen}, which states that for a real convex function $f$ and normalized weights $p_{k}$, $ f(\sum_{k}p_{k}x_{k})\leq \sum_{k}p_{k}f( x_{k})$. We now minimize the Lagrange $\Gamma$ function and obtain the following entanglement criterion:
\begin{subequations}
\begin{align}
\langle S_{z}\rangle &  =  \frac{1}{2}\left(\sum\limits_{\mu\leq\eta_{i}}\mu+\sum\limits_{\mu>\eta_{i}}\eta_{i}\right) \label{SepECS}\\
(\Delta S_{x})^{2} & \geq \frac{1}{4}\left(\sum\limits_{\mu\leq\eta_{i}}\mu^{2}+\sum\limits_{\mu>\eta_{i}}\eta_{i}^{2}\right).    \label{SepECV}
\end{align}
\end{subequations}
The summations in the above two equations are to be taken over $\left\lbrace\eta_{i}\right\rbrace_{i=1}^{N}$ depending on whether they are smaller or bigger than $\mu$, as indicated explicitly. We emphasize that the found minimum is tight in the sense that for every $\mu$, there is an associated state of the form given in Eq. \eqref{ExSepSt}, for which we get the lowest possible variance $(\Delta S_{x})^{2}$ associated with the corresponding value of $\langle S_{z}\rangle$. This minimum is obtained by having the weakly probed particles fully aligned along the $z$-axis, whereas the more strongly coupled particles have their spins rotated away from this direction, such that they point partially along the $\pm$ $x$-axis. As a special case of this result, we consider $\eta_{i}=1$ for all $i$. Eqs.~(\ref{SepECS}) and (\ref{SepECV}) then give $\langle S_{z}\rangle = \mu N/2$, $(\Delta S_{x})^{2} = \mu^{2}N/4$, and we recover the standard quadratic curve $(\Delta S_{x})^{2} \geq \langle S_{z}\rangle^{2}/N$ used to define the entanglement criterion $\xi^{2}<1$ with symmetric probing \cite{Anders}.

It is imperative to underline the importance of the coefficients $\eta_{i}$. Even though Eqs.~(\ref{SepECS}) and (\ref{SepECV}) give the entanglement criterion we were aiming for, it strongly depends on the distribution of $\eta_{i}$. This results in two important consequences: firstly, as $\eta_{i}$ depends on the experimental setup, we cannot define a single squeezing parameter which applies to all experiments. Secondly, we need to know each of these coefficients or, alternatively, their probability distribution $p(\eta)$ to evaluate the entanglement criterion. Notice that if the positions of the particles are fluctuating in time, their contribution to the measurement will be proportional to the average of the ensemble realizations. Hence Eqs. \eqref{SepECS} and \eqref{SepECV} should be replaced by their average values, which can be determined from the  probability distribution function of the coefficients. This in turn means that, as far as $p(\eta)$ remains unchanged during the measurement, fluctuations do not represent a problem for our entanglement criteria.

\begin{figure}[htbp]
	\centering
	\includegraphics[width=8 cm]{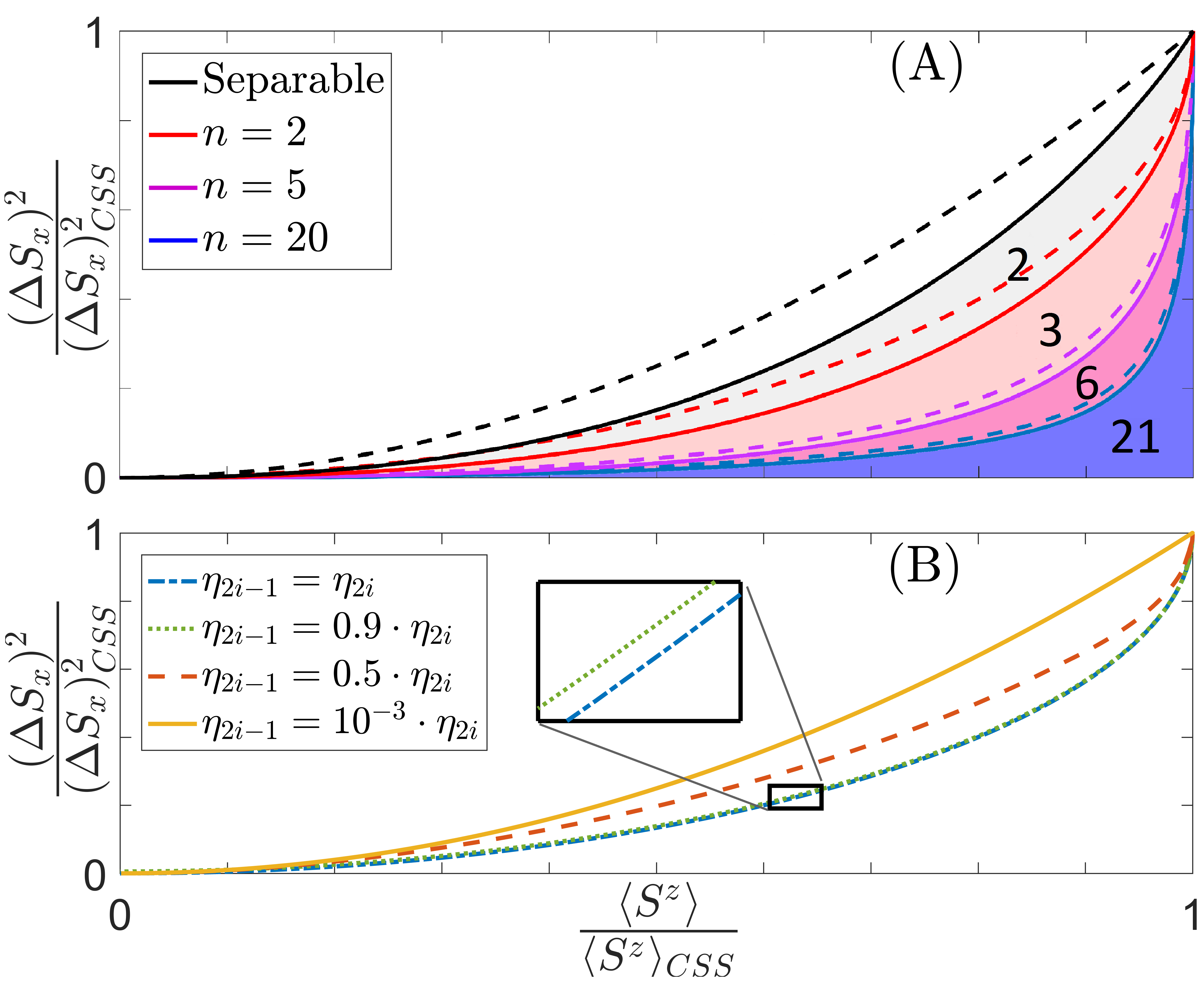}
	\caption[NewFig]{\textbf{(A)} Entanglement criteria for symmetric (dashed) and asymmetric (solid) probing for an atomic ensemble in a cylinder probed by a narrow Gaussian laser beam ($\nu=0.3$). The black, red, magenta and blue (listed from the top) curves correspond, respectively, to the minimum possible variance when allowing for separable, $n=2$, $5$, and $20$ particle-entangled states. The numbers between the lines therefore express the smallest possible group of entangled atoms in the corresponding regions. \textbf{(B)} Lower bounds for the normalized variance with respect to the average spin for a pair of atoms. The different curves correspond to different values of the coefficients $\eta_{2i-1}$ and $\eta_{2i}$, and were determined using a general wave function $\lvert\psi\rangle=C_{\uparrow\uparrow}\lvert \uparrow \uparrow\rangle+C_{\uparrow\downarrow}\lvert \uparrow \downarrow\rangle +C_{\downarrow\uparrow}\lvert \downarrow \uparrow\rangle+C_{\downarrow\downarrow}\lvert \downarrow \downarrow\rangle$. Along with $\eta_{2i-1}$ and $\eta_{2i}$, $\lvert\psi\rangle$ fully specify the state of the system; we can therefore use the complex numbers $C_{kl}$ ($k,l=\uparrow$ or $\downarrow$) as variables in the Lagrange $\Gamma$ function, and obtain the minimal noise for each value of the mean spin, and $\eta_1$ and $\eta_2$. Letting the fraction $\eta_{2i-1}/\eta_{2i}$ vary continuously in the interval $[0,1]$, we find that the minimal noise is obtained for $\eta_1=\eta_2$.}
	\label{NewFig}
\end{figure}

For illustrating how Eqs.~(\ref{SepECS}) and (\ref{SepECV}) can be used as an entanglement criterion, we introduce a simple example for $p(\eta)$. Let us consider an atomic ensemble in a cylinder with radius $R$. A Gaussian probing beam centred in the middle interrogates the atoms along the axis of the cylinder. We assume that the particles are uniformly distributed, so that we have the probing coefficient $\eta(x) = e^{-\frac{x^{2}}{\sigma^{2}}}$, where $x$ is the distance from the axis of the laser. In the interval $\eta\in[e^{-\frac{1}{\nu^{2}}},1]$ we have $p(\eta)=\nu^2/\eta$, where $\nu=\sigma/R$ is the only parameter characterizing such a model. Inserting this into Eqs.~(\ref{SepECS}) and (\ref{SepECV}) we find the entanglement criterion, shown by the solid black line (first solid curve from the top) in Fig. \ref{NewFig} (A). This curve represents the lowest variance one can attain for a separable state, and thus if a variance lower than this is measured for the ensemble, it signifies that (at least) bi-partite entanglement is present in the system. Note that the mean spin and the variance are normalized to the value obtained for the $\lvert \text{CSS}\rangle$. This corresponds to the procedure typically employed in the experiments \cite{Exp2,Exp3,Thompson}.
As evident from the plot, the dashed black line which is obtained for symmetric probing of the atomic ensemble significantly differs from the present case.

To better understand the dependence on the parameter $\nu$, we investigate the behavior in the extreme case $\nu\ll 1$, where the probe is much narrower than the cylinder. Let us first consider the limit $\langle S_{z}\rangle/\langle S_{z}\rangle_{\text{CSS}}\rightarrow 1$. In this case the slopes of the symmetric and asymmetric criteria will differ by a factor of two. This means that in this limit twice as much squeezing is required to claim entanglement. In the other extreme $\langle S_{z}\rangle/\langle S_{z}\rangle_{\text{CSS}}\rightarrow 0$, the ratio between the minimal noise in the symmetric and asymmetric case for a given $\langle S_{z}\rangle$ scales as $1/(2\nu^2)$, and can thus be arbitrarily large for $\nu\rightarrow 0$ (Note that this limit also implies $N\rightarrow \infty$, and hence this does not contradict the lower limit of $1/N$ derived above). This clearly shows the inadequacy of the squeezing parameter $\xi^{2}$ as an entanglement criterion in the case of asymmetric probing. The present example is motivated by the specific physical systems of atomic ensembles probed by Gaussian laser beams. However, spin squeezing is used in other physical systems \cite{SimilExp1,SimilExp2,SimilExp3} as well, where the distribution may be even more asymmetric, resulting in a larger difference between the criteria.

Above  we have established a criterion for showing the existence of entanglement. We now extend this to multi-partite entanglement by allowing the particles to be entangled in groups of $n$ atoms. Thus, the violation of the minimum obtainable variance for $n$-particles will identify samples with at least $(n+1)$ entangled atoms. Assuming for simplicity that the total number of particles fulfil $N=l n$ for some integer $l$, we define the density matrix containing at most $n$ entangled particles to be given by 
\begin{equation}
\rho =\sum\limits_{k}p_{k}\bigotimes_{i=1}^{N/n}\rho^{(k)}_{[n(i-1)+1],...,ni}\label{rho}.
\end{equation}
As before, the $p_{k}$'s represent probabilities, and $\rho^{(k)}_{n(i-1)+1,...,ni}$ is the density matrix of $n$ entangled particles for any $k$, $i=1,...,N/n$. In general, we should allow for permutations between the atoms, but this does not change the result \cite{MiaTesi}. Moreover, generalizations to non integer $N/n$ can be done as in Ref. \cite{EntSque4}. As before, we evaluate the average spin $\langle S_{z}\rangle$ and the variance $(\Delta S_{{x}})^{2}$ for the density matrix in Eq. \eqref{rho}. We then minimize the Lagrange function to find the $(n+1)$ particle entanglement criterion.

We start by considering the simplest case of $n=2$ entangled atoms, for which $\rho =\bigotimes_{i=1}^{N/2}\rho_{2i-1,2i}$. Generalizations to any $n$ can be done with minor adjustments which we specify later. For clarity, we have omitted the summation over the index $k$, that would in the end disappear from the equivalent of Eqs. \eqref{2PS} and \eqref{2PV}. Using this density matrix, it is possible to obtain $\langle S_{z}\rangle = \sum_{i}\langle S_{z}\rangle_{2i-1,2i}$ and $(\Delta S_{x})^{2} \geq  \sum_{i}\left(\Delta S_{x}\right)^{2}_{2i-1,2i}$. Therefore a first expression for the minimization is given by
\begin{equation}
\begin{split}
\Gamma = & \sum\limits_{i=1}^{\frac{N}{2}}\Bigg\lbrace\left[\frac{\eta_{2i-1}^{2}+\eta_{2i}^{2}}{4}\right]\frac{(\Delta S_{x})^{2}_{2i-1,2i}}{(\Delta S_{x})^{2}_{\text{CSS};2i-1,2i}} + \\ &-\mu \left[\frac{\eta_{2i-1}+\eta_{2i}}{2}\right]\frac{\langle S_{z}\rangle_{2i-1,2i}}{\langle S_{z}\rangle_{\text{CSS};2i-1,2i}} \Bigg\rbrace, \label{Gamma2Part}
\end{split}
\end{equation}
where we have written it by renormalizing with respect to the state $\lvert \text{CSS} \rangle$, using $(\Delta S_{x})^{2}_{\text{CSS};2i-1,2i}=(\eta_{2i-1}^{2}+\eta_{2i}^{2})/4$ and $\langle S_{z}\rangle_{\text{CSS};2i-1,2i} = (\eta_{2i-1}+\eta_{2i})/2$. In Eq. \eqref{Gamma2Part}, all contributions in the sum refer to two (possibly entangled) particles only. This means that by minimizing all of them independently, we obtain a global limit for the whole state. Let us therefore consider the contribution of a single pair described by the coefficients $\eta_{2i-1}$ and $\eta_{2i}$. The minimum that the normalized variance $(\Delta S_{x})^{2}_{2i-1,2i}/(\Delta S_{x})^{2}_{\text{CSS};2i-1,2i}$ can achieve for a given value of the mean spin $\langle S_{z}\rangle_{2i-1,2i}/\langle S_{z}\rangle_{\text{CSS};2i-1,2i}$ depends on the ratio $\eta_{2i-1}/\eta_{2i}$. By minimizing the noise numerically we find this minimum, shown in Fig. \ref{NewFig} (B). As seen in the plot, the minimum is achieved when the coefficients are equal: $\eta_{2i-1}=\eta_{2i}$. This in turn means that the corresponding minimization curve is already known \cite{Anders}
\begin{equation}
\frac{\left(\Delta S_{{x}}\right)_{2i-1,2i}^{2}}{\left(\Delta S_{{x}}\right)_{\text{CSS};2i-1,2i}^{2}} \geq 1-\sqrt{1-\left(\frac{\langle S_{z}\rangle_{2i-1,2i}}{\langle S_{z}\rangle_{\text{CSS};2i-1,2i}}\right)^{2}}\label{2PLB}.
\end{equation}
Inserting Eq.~(\ref{2PLB}) into Eq.~(\ref{Gamma2Part}), and using that the minimum is found for the coefficients $\eta_{i}$ being pairwise equal, we can rewrite the Lagrange $\Gamma$ function for pairs of entangled atoms and minimize it for a given $\mu$. In this way we find that
\begin{subequations}
	\small
	\begin{align}
	\frac{\langle S_{z}\rangle}{\langle S_{z}\rangle_\text{CSS}} &  =  \left(\sum\limits_{i=1}^{N/2}\eta_{i}\right)^{-1}\sum\limits_{j=1}^{N/2}\mu\frac{2\eta_{j}}{\sqrt{\eta_{j}^{2}+4\mu^{2}}}\label{2PS}\\
	\frac{(\Delta S_{x})^{2}}{(\Delta S_{x})_\text{CSS}^{2}} & \geq \left( \sum\limits_{i=1}^{N/2}\eta_{i}^{2} \right)^{-1} \sum\limits_{j=1}^{N/2}\eta_{j}^{2}\left[ 1-\frac{\eta_{j}}{\sqrt{\eta_{j}^{2}+4\mu^{2}}} \right]\label{2PV}
	\end{align}
\end{subequations}
is a lower bound on $(\Delta S_{x})^{2}/(\Delta S_{x})^{2}_\text{CSS}$ for the given $\langle S_{z}\rangle/\langle S_{z}\rangle_\text{CSS}$, assuming that the particles are only entangled in pairs. Hence the violation of this bound is a sufficient criterion for proving tri-partite entanglement. For the model introduced above, we plot this criterion as the red solid line (second solid curve from the top) in Fig.\ref{NewFig} (A). Measurement results below this line will therefore signify samples with at least three particle entanglement. The red dashed curve corresponds to the case of symmetric probing of the ensemble. As for the separable case, the difference between them is evident.

For higher multi-partite entanglement ($n > 2$), the procedure for obtaining the criteria is similar. The only differences reside in finding the condition on $\eta_{i}$ for minimizing the Lagrange function, and the analogue of Eq. \eqref{2PLB} for $n>2$. To do this, we need to determine the combination of coefficients $\left\lbrace\eta_1,...,\eta_n\right\rbrace$ which minimize the Lagrange function for $n$ particles. For small $n$ this can be done numerically by minimizing the Lagrange function. The complexity of the minimization, however, grows exponentially with $n$ making the problem intractable for large values. We have explicitly done the minimization for all $n\leq 8$. We found that the minimum is always obtained for the coefficients being equal, similar to the $n=2$ case. Assuming this to be a general property valid for all $n$, we can seek the equivalent of Eq. (\ref{2PLB}). For $n>2$ a tight lower bound on the variance with respect to the average spin is not known. However, since we consider all $\eta_i$ to be the same, we can use the results of Ref. \cite{Anders}, which give two possible methods. The first one is to exploit the lower bound:
\begin{widetext}
	\begin{equation}
	\frac{(\Delta S_{x})^{2}}{(\Delta S_{x})_\text{CSS}^{2}}\geq 1 + \frac{n}{2}\left[ 1 - \frac{\langle S_{z}\rangle^{2}}{\langle S_{z}\rangle_\text{CSS}^{2}}\right] - \sqrt{\left[ 1 + \frac{n}{2}\left( 1-\frac{\langle S_{z}\rangle^{2} }{\langle S_{z}\rangle_\text{CSS}^{2} } \right) \right]^{2} - \frac{\langle S_{z}\rangle^{2} }{\langle S_{z}\rangle_\text{CSS}^{2} } }.\label{kPLB}
	\end{equation}
\end{widetext}
This condition is not tight since in general one cannot find a state which reaches the bound. For large $n$, however, the bound in Eq. \eqref{kPLB} is close to the true minimum. An alternative method is to use a numerical procedure, which is efficient for any even $n$ and represents a tight lower bound \cite{Anders}. With these methods it is possible to rewrite the equivalent of Eq. \eqref{Gamma2Part} for $n$ entangled particles, as a function made of $l=N/n$ $n$-uplets contributions. These, in a similar way to the $n=2$ case, can be minimized one by one in order to find all the other multi-partite entanglement criteria, i.e. the equivalent of equations (\ref{2PS}) and (\ref{2PV}) for $n>2$.  Examples of these criteria are shown in Fig.\ref{NewFig}, (A) for the simple model described above. The solid curves correspond to the lower bounds for the variance that states with $n = 5$ (magenta, third solid line from the top) and $n = 20$ (blue, bottom) entangled atoms cannot go below. The first one is obtained through Eq. \eqref{kPLB}, the latter using the mentioned numerical procedure.
%The dashed curves for $n = 5$ (magenta) and $n = 20$ (blue) in the figure corresponds to the entanglement criteria one gets for the atoms homogeneously coupled to light.

\begin{figure}[h!tbp]
	\centering
	\includegraphics[width=9 cm]{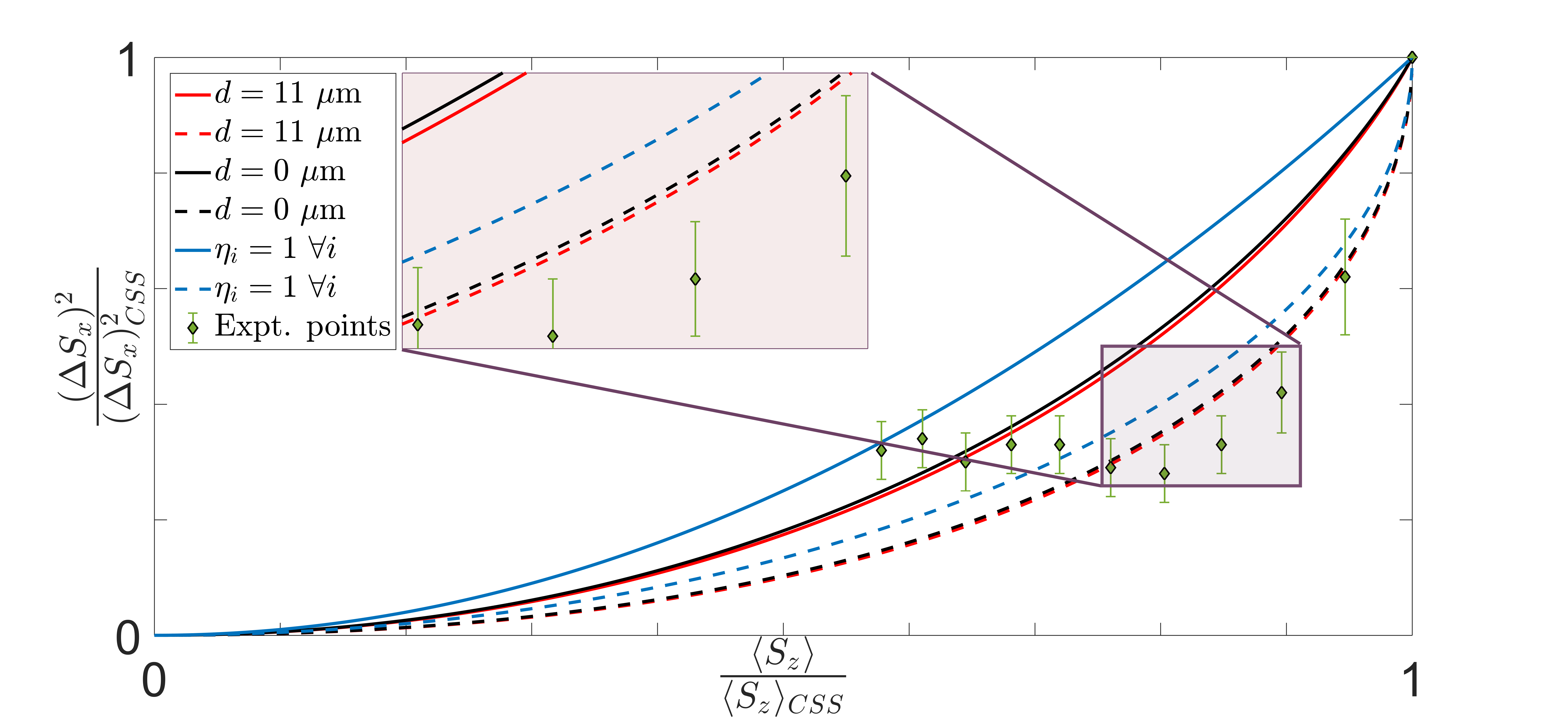}
	\caption[Figure 2]{Entanglement and $3$-particle entanglement criteria for the experiment described in Ref. \cite{Appel}. The solid curves give bounds on the variance assuming a separable state, while the dashed curves give the bound when allowing for bi-partite entanglement. The colours blue (top), black (middle) and red (bottom) correspond respectively to symmetric probing, asymmetric probing with no displacement and asymmetric probing with displacement between the FORT and the probes.
		Parameters used are: $T = 50$ $\mu$K, $\omega_{t} = 50$ $\mu$m, $\lambda_{t} = 1032$ nm, $\omega_{p} = 27$ $\mu$m, $\lambda_{p} = 852$ nm, and $V_{0}/k_{B}=1.73\cdot 10^{-4}$K.  }
	\label{Fig2}
\end{figure}

As an application, we revisit the experimental results of Ref. \cite{Appel}, and utilize our criteria to verify entanglement of the atoms. In the experiment, an ensemble of $\gtrsim 10^{5}$ Cs atoms was cooled and trapped in a Far Off Resonant Trap (FORT). These atoms were then probed with a two colour quantum non demolition scheme \cite{QND2,QND3,QND4,QND5,QND6}, that allows measuring the variance and average spin. For applying our entanglement criteria we need to know the atomic distribution in the ensemble and the resulting statistics $p(\eta)$ of the coefficients $\eta_{i}$. We assume a thermal distribution of the atoms in the ensemble of the form $\rho(r,z) \propto re^{-\frac{V(r,z)}{k_{b}T}}$ \cite{DanielThesis}. Here, $T$ is the temperature, $r$ represents the distance from the beam's axis, $k_{b}$ is the Boltzmann constant and $z$ is the longitudinal position. The potential $V(r,z)$ generated by the Gaussian FORT beam is of the form $V(r,z) = V_{0}\frac{\omega_{t}^{2}}{\Omega_{t}(z)^{2}}\exp\left[-\frac{2r^{2}}{\Omega_{t}(z)^{2}}\right]$. In this equation, $V_{0}$ is the minimum of the potential, $\omega_{t}$ the waist of the trapping beam, and $\Omega_{t}(z)=\omega_{t}\sqrt{1+\left(\frac{z\lambda_{t}}{\pi \omega_{t}^{2}}\right)^{2}}$ the spatially varying spot size of the laser with wavelength $\lambda_{t}$. In the experiment the sample was well described by a thermal distribution in the radial direction $r$ only, not in the longitudinal direction $z$, since the atoms are loaded into the FORT at a well defined position and do not have time to expand to the full length of the system. Thereby, the entanglement criteria we derive with a thermal distribution are lower than the real ones, making their violation harder. To find the coefficients $\eta_{i}$ we use that the intensity of the probe in the experiment was described by a Gaussian profile \cite{Appel} 
\begin{equation}
\eta(r,z) = \frac{\omega_{p}^{2}}{\Omega_{p}(z)^{2}}e^{-\frac{2r^{2}}{\Omega_{p}(z)^{2}}}.
\end{equation}
The parameters $\omega_{p}$ and $\Omega_{p}(z)$ are the waist and spot size of the probing beam, and satisfy a similar relation as the trap parameters $\omega_{t}$ and $\Omega_{t}$, except that the relevant wavelength here is $\lambda_{p}$. We can now evaluate the coefficients $\eta_{i}$ \cite{MiaTesi} according to their probability distribution, and find the multi-partite entanglement criteria. This is shown in Fig. \ref{Fig2}, where we also include the experimental results of Ref. \cite{Appel}. All of the required parameters except the temperature can be directly derived from the experiment \cite{Appel}. To be conservative, we chose a value $T= 50 \text{ $\mu K$}$, that is slightly higher than typical values in similar setups \cite{DanielThesis}. This gives a more asymmetric  distribution of the coefficients $\eta_{i}$ and thereby a more strict limit. Furthermore, we consider a possible displacement $d$ between the probe and the FORT. Such a misalignment between the probing and the trapping beams again results in lowering of the curve, making it harder to detect entanglement. In the experiment the overlap between the probe and the FORT was set by optimizing the detected signal from the atoms. We therefore assign an upper limit to the displacement of $d=11$ $\mu$m, set by the requirement that the signal is at least $90\%$ of the maximal one.
As seen in Fig. \ref{Fig2} and highlighted in the inset, some of the experimental points lie below the dashed (red \& black, lowest) curves. Since these curves correspond to the lower limits obtained by allowing bi-partite entanglement, this signifies the presence of tri-partite entanglement among the atoms in the experiment.

In summary, we have considered the problem of verifying entanglement for spin systems subject to  asymmetric probing. In this case a naive generalization of the spin squeezing entanglement criteria --  derived for symmetric probing -- cannot be used to verify entanglement in the system. We have explicitly derived
new criteria for bi-partite and multi-partite entanglement. A key feature of our procedure is that it can easily be adapted to any setup, once we know the probability distribution of the coupling coefficients. We have demonstrated this by applying it to the experimental data of Ref. \cite{Appel}, proving tri-partite entanglement among the atoms in the ensemble. Our criteria thus provide an effective means of detecting the degree of entanglement in spin squeezing experiments like atomic ensembles. The procedure is, however, equally applicable to other physical systems \cite{Squuez2,SimilExp1,SimilExp2,SimilExp3}. Moreover, since the improvement of atomic clocks is related to the spin squeezing parameter, our results suggest that there exist unentangled states which would give a better performance of atomic clocks than $\lvert\text{CSS}\rangle$. 
\begin{acknowledgements}
We gratefully acknowledge funding from the European Union Seventh Framework Programme through the ERC Grant QIOS and INTERFACE, and the Danish Council for Independent Research (DFF). We thank J\"{o}rg Helge M\"{u}ller and Eugene Polzik for fruitful discussions.
\end{acknowledgements}

\end{document}